\documentclass[twocolumn,prd,nofootinbib,aps,floats,floatfix,amsmath,amssymb,secnumarabic]{revtex4} %

\usepackage{graphicx}
\usepackage[usenames,dvipsnames]{color}
\usepackage{amsmath,amssymb}
\usepackage{slashed}
\usepackage[colorlinks,citecolor=blue]{hyperref}

\def\sfrac#1#2{{\textstyle{#1\over #2}}}
\newcommand{\phm}{\phantom{-}}
\newcommand{\be}{\begin{equation}}
\newcommand{\ee}{\end{equation}}
\newcommand{\ba}{\begin{array}}
\newcommand{\ea}{\end{array}}
\newcommand{\bea}{\begin{eqnarray}}
\newcommand{\eea}{\end{eqnarray}}

\newcommand{\W}{{\scriptscriptstyle W}}
\newcommand{\Z}{{\scriptscriptstyle Z}}
\newcommand{\R}{{\scriptscriptstyle R}}

\newcommand{\SSS}{{\scriptscriptstyle S}}
\newcommand{\T}{{\scriptscriptstyle T}}

\newcommand{\comment}[1]{}

\begin{document}
\title{Composite magnetic dark matter and the 130 GeV line}
\author{James M.\ Cline}
\email{jcline@physics.mcgill.ca}
\affiliation{Department of Physics, McGill University,
3600 rue University, Montr\'eal, Qu\'ebec, Canada H3A 2T8}
\author{Andrew R.~Frey}
\email{a.frey@uwinnipeg.ca}
\affiliation{Dept.~of Physics and Winnipeg Institute for Theoretical
Physics, University of Winnipeg, Winnipeg, MB, R3B 2E9, Canada} 
\author{Guy D.\ Moore}
\email{guymoore@physics.mcgill.ca}
\affiliation{Department of Physics, McGill University,
3600 rue University, Montr\'eal, Qu\'ebec, Canada H3A 2T8}

\begin{abstract}

We propose an economical model to explain the apparent 130 GeV gamma
ray peak, found in the Fermi/LAT data, in terms of dark matter (DM)
annihilation through a dipole moment interaction. 
The annihilating dark matter particles represent a
subdominant component, with mass density $7-17$\% of the total DM
density; and they only annihilate into $\gamma\gamma$, $\gamma Z$, and
$ZZ$, through a magnetic (or electric) dipole moment. Annihilation
into other standard model particles is suppressed, due to a DM mass
splitting in the magnetic dipole case, or to $p$-wave scattering
in the electric dipole case. In either case, the observed signal
requires a dipole moment of strength  $\mu\sim 2/$TeV. We argue that
composite models are the preferred means of generating such a large
dipole moment, and that the magnetic case is more natural than the
electric one.  We present a simple model involving a scalar and
fermionic techniquark of a confining SU(2) gauge symmetry.  We point
out some generic challenges for getting such a model to work.
The new physics leading to a sufficiently large dipole moment is below the TeV
scale, indicating that the magnetic moment is not a valid effective
operator for LHC physics, and that production of the 
strongly interacting constituents, followed by techni-hadronization,
is a more likely signature than monophoton events.
In particular, 4-photon events from the decays of bound state pairs
are predicted.

\end{abstract}
\maketitle


In the past few months a number of studies of publicly available Fermi
Large Area Telescope \cite{Atwood:2009ez}  data have emerged that find
evidence for a spectral peak of energy $\sim 130$ GeV from the
galactic center  \cite{Bringmann:2012vr,Weniger:2012tx,Tempel:2012ey},
possibly accompanied by a second peak with lower energy $\sim 116$ GeV
\cite{Su:2012ft,Rajaraman:2012db}.  Ref.\ \cite{Hektor} moreover finds
corroborating evidence from outside the galaxy, originating from
several galactic clusters, and ref.\ \cite{Su:2012zg} sees marginal evidence for
sources unassociated with known structures, though this has been questioned
\cite{Hooper:2012qc,Mirabal:2012za,Hektor:2012jc}.  The morphology of the signal from the galactic 
center is consistent
with dark matter annihilating into monoenergetic photons, but not with
dark matter decays \cite{Buchmuller:2012rc, Hektor}.  In terms of the
thermally averaged cross section required for the standard DM relic
density, $\langle\sigma v\rangle_0= 3\times 10^{-26}$cm$^3$/s,  the
annihilating DM hypothesis is consistent with a cross section
estimated to be $0.04$  $\langle\sigma v\rangle_0$
\cite{Weniger:2012tx,Buchmuller:2012rc}  or $0.1 \langle\sigma
v\rangle_0$ \cite{Tempel:2012ey,Hektor}.  Fermi/LAT itself places a
limit of $\langle\sigma v\rangle_{\gamma\gamma} < 0.035\langle\sigma
v\rangle_0$  on this cross section (for an Einasto profile)
\cite{Goodman:2010qn,Ackermann:2012qk}, marginally below the required value. 
Searches for the signal from  dwarf galaxies yield the weaker limit 
$\langle\sigma v\rangle_{\gamma\gamma} < 1.3\langle\sigma
v\rangle_0$ \cite{GeringerSameth:2012sr}.  If the 130 GeV line is accompanied by a gamma continuum
due to additional annihilation channels into charged particles, the
limit on the cross section for these processes is estimated
by ref.\ \cite{Buckley:2012ws} to be 
$\langle\sigma v\rangle_{\rm other} < 3.5 \langle\sigma
v\rangle_{\gamma\gamma}$ for the $b\bar b$ channel, while
ref.\ \cite{Cohen:2012me} obtains the weaker limit		
$\langle\sigma v\rangle_{\rm other} < 20 \langle\sigma
v\rangle_{\gamma\gamma}$.  (Weaker limits are found in ref.\
\cite{Cholis:2012fb}, and stronger ones in \cite{Huang:2012yf}). While the latter is only marginally
in conflict with the DM getting its standard relic density from annhilating primarily
into charged particles, the former would definitely forbid this
scenario, presenting a challenge to many models.

Although some authors have expressed skepticism about the DM
interpretation  of the 130 GeV  line
\cite{Profumo:2012tr,Boyarsky:2012ca}, there has been a great deal of
interest in this possibility both from the experimental analysis
perpsective  \cite{Ibarra:2012dw}-\cite{Frandsen:2012db} and in 
theoretical model-building \cite{Dudas:2012pb}-\cite{Park:2012xq}.
(Ref.\ \cite{Jackson:2009kg}, which preceded the observation of the
130 GeV line, is also a competing theory.)   In the present paper, we
consider a distinctive model similar to those discussed in refs.\ 
\cite{Weiner:2012cb,Tulin:2012uq}, in which the dark matter interacts primarily
through a large magnetic dipole moment (MDM).  
Direct detection of dipolar dark matter has
been the focus of many recent papers, dealing either with
elastic  \cite{Sigurdson:2004zp}-\cite{Fitzpatrick:2010br} or 
inelastic models \cite{Profumo:2006im}-\cite{Patra:2011aa}
(for recent work on indirect detection, see \cite{Heo:2012dk}).  
Here we show that Majorana DM with a transition MDM and 
a relatively large mass splitting can economically explain
the observed  gamma ray line, in particular  if we relax
the usual assumption that the candidate particle must be the
dominant form of dark matter. (As we will discuss, 
DM with an electric dipole moment
(EDM) and a smaller mass splitting can also work, though we do not
favor this scenario.)   Moreover we remove the tension of too
large a $\gamma$-ray continuum by arranging for the $2\gamma$ annihilation
channel (along with the accompanying $\gamma Z$ and $ZZ$ channels) 
to be the dominant one.  Majorana DM $\chi_1$ can have a
transition magnetic moment to a heavier state $\chi_2$,
\be
	\sfrac12{\mu_{12}}\,\bar\chi_1\sigma_{\mu\nu}\chi_2 F^{\mu\nu} \,.
\label{muint}
\ee
It is natural to assume that the interaction (\ref{muint})
came originally from standard model hypercharge, and is therefore 
accompanied by a corresponding coupling to the $Z$ boson field 
strength.

\begin{figure}[tb]
\centering
\includegraphics[width=0.5\textwidth]{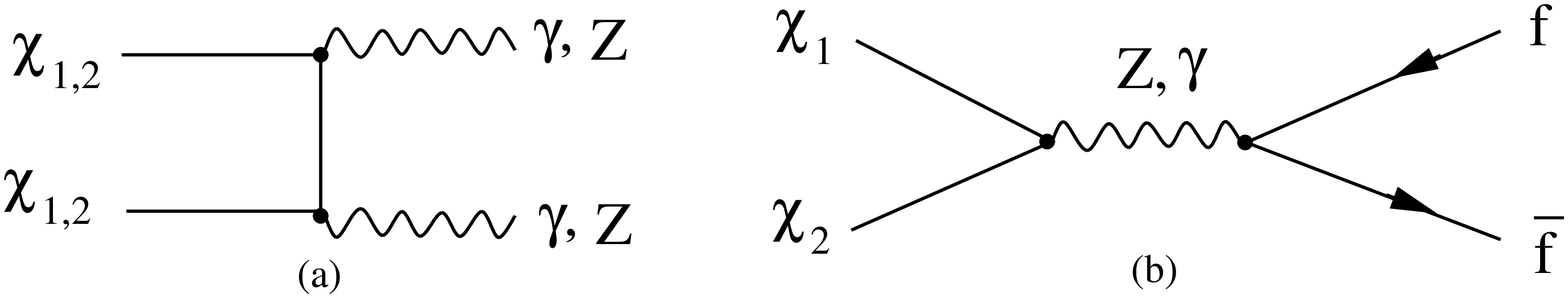}
\caption{Annihilation processes for determining the relic density.
(a) Annihilations $\chi_{1,2}\chi_{1,2}\to \gamma\gamma$ (the
$u$-channel diagram with crossed initial or final states is not shown); (b)
coannihilations $\chi_{1}\chi_{2}\to f\bar f$ where $f$ is a charged
SM particle (or Higgs in the case of $Z$ intermediate state).}
\label{fig1}
\end{figure}

With the interaction (\ref{muint}), there exist both annihilations
 $\chi_{1,2}\,\chi_{1,2}\to \gamma\gamma,\gamma Z, ZZ$,  
as well as  coannihilations
$\chi_1\chi_2\to f\bar f$, where $f$ is any charged standard model
particle, or $hZ$ in the case of the $Z$ intermediate state.  These are pictured
in fig.\ \ref{fig1}.  As was shown in ref.\ \cite{Tulin:2012uq}, 
the coannihilations
suppress the relic density below its required value;
but a mass splitting $m_{\chi_2}-m_{\chi_1}\gtrsim 10$ GeV
suppresses the coannihilations enough to get the right
relic density.  Therefore a rather large mass splitting is a needed
ingredient in the magnetic dipole case.  Such a large mass splitting of
course makes direct detection of $\chi_1$ from scattering on nuclei
impossible,%
\footnote{at first order in perturbation theory; integrating out
the excited DM state of course leads to a small elastic transition
at higher order \cite{Batell:2009vb}.}
because of the large inelasticity \cite{TuckerSmith:2001hy}.
On the other hand, if the coupling is via an
electric dipole moment, which has an extra factor of $\gamma_5$, 
then the troublesome annihilations into
SM particles are $p$-wave suppressed and can be ignored.  

In any case, there
is value of $\mu$ such that the observed line intensity is
achieved, allowing for the relic density to be smaller (by a factor of  $\sim
10$) than the standard value, and then $\chi$ is a subdominant component
of the total DM.  This possibility was noticed, though explored in more
detail for the case that co-annihilations determine the relic
density, in ref.\ \cite{Weiner:2012cb}.
We do not concern ourselves here with the identity of the
majority of dark matter, although it is interesting to note that it could be
for example a $\sim 10$ GeV particle as suggested by observations of the
CoGeNT and CRESST experiments.  

We will assume that the DM carries no SU(2)$_L$ 
charge, but only weak hypercharge.  In that case the interaction
(\ref{muint}) is accompanied by the analogous coupling to the
$Z$ boson field strength, suppressed by $\tan\theta_\W \equiv t_\W$:
\be
- \sfrac12{t_\W\mu_{12}}\,\bar\chi_1\sigma_{\mu\nu}\gamma_5\chi_2
        Z^{\mu\nu} \,.
\label{Zmuint}
\ee
The relative sign between (\ref{muint}) and (\ref{Zmuint}) comes from
the relation between hypercharge and the mass eigenstates, $B^\mu
=c_\W A^\mu -s_\W Z^\mu$.

\def\sigvv{\langle \sigma_{\gamma\gamma} v\rangle}
\def\sigfull{\langle \sigma v\rangle}
\def\sigo{\langle \sigma v\rangle_0}
\bigskip\bigskip\bigskip \noindent
{\bf Determining $\mu$ and relic abundance.}  
We assume that the dominant annihilation channels are
to $\gamma\gamma,\gamma Z, ZZ$ via a dipole moment, as shown in
fig. \ref{fig1}(a)---both in the context of current galactic dark
matter signals, as well as that of the annihilations which establish
the relic abundance.\footnote{This is a more detailed discussion of a 
comment made in \cite{Weiner:2012cb} for DM with a Rayleigh interaction.}  
Let us call the cross-section to two photons
$\sigvv$; then using Eq.~(\ref{Zmuint}),
the total cross-section is 
$\sigfull \simeq \sigvv/
\cos^4\theta_\W$ (neglecting corrections from the different kinematics of
a $Z$ final state, see appendix \ref{ann}).  We also write the density
of 130 GeV dark matter as
$n$, which will be smaller than the standard relic density $n_0$.
The total 130 GeV photon production rate scales as
$\sigvv n^2$.
According to refs.\ \cite{Weniger:2012tx,Tempel:2012ey} the observed
$\gamma$ ray signal corresponds to
\be
\sigvv n^2 = (0.04\mbox{--}0.1)
\sigo n_0^2 \,,
\ee
where $\sigo$ is the cross-section which would lead,
via the usual freezeout calculation, to a relic density of $n_0$.
Since the relic density is approximately inversely proportional to the
annihilation cross-section,
\be
n \simeq \frac{\sigo}{\sigfull}
n_0 \,,
\ee
the photon producing cross-section should satisfy
\be
\frac{\sigvv}{\sigo} \simeq \frac{\cos^8 \theta_\W}{(0.04\mbox{--}0.1)} \,.
\ee
Accounting for the dependence on the mass splittings given
in appendix B, we find that the
dipole moment should lie in the range
\be
1.6\,f(r)\ \mbox{TeV}^{-1} < \mu_{12} < 2.0\,f(r)\ \mbox{TeV}^{-1}
\label{edm_range}
\ee
where $r=m_{\chi_2}/m_{\chi_1}$ and $f=\sqrt{(1+r^2)/2r}$.
The corresponding fractional relic abundance is
\be
{0.07\over f(r)} < {n\over n_0} < {0.17\over f(r)} \,.
\ee
Notice that a larger dipole moment leads to more complete annihilation during
freezeout, and hence to a smaller abundance and a smaller
photon flux.

If another annihilation mechanism was active in the early Universe---such as coannihilation, which can occur if the $\chi_1$ to $\chi_2$ mass
splitting is not too large \cite{Tulin:2012uq}---then $\sigvv$ and
hence $\mu_{12}$ need to be smaller, and the relic density fraction is
larger.  In the extreme case that coannihilation controls the relic
density and $n=n_0$, then $\sigvv$ can be about 100 times smaller, and
hence $\mu_{12}$ can be 10 times smaller.  The coannihilation cross section
is just large enough to give the right relic density (if the mass
splitting is small) for this smaller value of $\mu_{12}$.  

\bigskip\noindent
{\bf Direct detection.}  If the DM candidate $\chi$ was Dirac, an elastic 
EDM or MDM as large as in Eq.~(\ref{edm_range}) would be ruled out by many
orders of magnitude, by
direct detection searches for elastic scattering on nuclei.
In ref.\ \cite{Chang:2010en} (see also
\cite{Profumo:2006im}-\cite{Masso:2009mu},\cite{Feldstein:2010su}-\cite{Patra:2011aa})
it was shown that by splitting the Dirac particle into two Majorana states
with mass splitting of $\sim 150$ keV, a magnetic moment $\sim 10^{-2}\mu_N
= 1.6/$TeV is compatible with direct detection limits (though somewhat too
large to make the DAMA annual modulation signal compatible
with these bounds).  In the case of MDM, scatterings can be dominated 
either by dipole-charge interactions (which lack the $1/v^2$ enhancement
present in the EDM case) or dipole-dipole interactions, depending on the
nucleus.
\comment{
Depending upon the nucleus, 
the scatterings can be dominated by dipole-dipole or dipole-charge
interactions between the DM and the nuclei, both of which
are velocity-suppressed, compared to EDM, in the case of MDM.}

The above determination was made assuming that the dipolar
DM had the standard relic density; in our case the density is lower by 
about a factor of 10.  Thus our DM candidate looks similar to DM with the
full relic density but a smaller magnetic moment, reduced by the factor
$\sim\sqrt{10}$.  This reduction by half an order of magnitude is enough
to be marginally compatible with fits to the DAMA data, with mass splitting
$\sim 125$ keV; see fig.\ 2 of \cite{Chang:2010en}.
However we cannot reconcile such a small mass 
splitting with the need to suppress the $\chi_1\chi_2 \to f\bar f$
coannihilation channels.  

On the other hand, for EDM dark matter, the coannihilations are
$p$-wave suppressed even if the mass splitting is small, which
suggests the possibility of combining the EDM explanation of the 130
GeV line with direct detectability.  EDM scattering is dominated by
interaction with the charge of the proton, and  is enhanced by $1/v^2$
relative to dipole-charge scattering of an MDM (see the erratum of ref.\
\cite{Sigurdson:2004zp}).  For inelastic scattering with mass
splitting $\delta M$, we find that this translates to an  enhancement
by the factor $m_n/\delta M$ in the cross section.  Therefore one
would have to reduce the EDM by a factor of $\sqrt{m_N/\delta M}\sim
100$ relative to the above estimate for it to be relevant for direct
detection, which would then make it too small to explain the 130 GeV
line. We find the same conclusion  whether dipole-dipole or
dipole-charge scattering dominates in the MDM interaction. (Notice that one cannot increase $\delta M$ much above 100 keV
for direct detection since the minimum required DM velocity would then
exceed the escape velocity of the galaxy.) 

\bigskip\noindent
{\bf Model requirements.}
Consider first that the dark matter is a neutral fundamental field
$\chi$.  One way it can obtain a large
dipole moment is through a Yukawa
interaction $y\bar\chi\phi S$  to heavy (hyper)charged states $\psi$,
$S$.  If all the states have masses
$\sim 100$ GeV, then an EDM of order $1/{\rm TeV}$ is only achieved by
pushing $y$ to the perturbative limit $\sqrt{4\pi}$ and 
tuning the particles in the loop to be nearly on shell
\cite{Sher:2002ij}.
Large MDMs require similar tuning of parameters.  Thus any
perturbative origin requires stretching the limits of perturbation
theory; probably the effective theory describing such states cannot be
valid much higher than the particle mass scales.  In fact, 
\cite{Weiner:2012gm} give a perturbative construction of a large MDM,
which requires large couplings, messenger masses not far above the DM mass,
and dimension 7 Rayleigh operators to explain the Fermi signal.

On the other hand, large magnetic moments are known to arise for neutral
composite particles.  The neutron is a good example, with
$\mu=-1.91(e/2m_p)$, which is approximately the sum
of the magnetic moments of the constituent quarks (treating them to have
constituent quark masses $\sim m_N/3$, see for
example ref.\ \cite{georgi}).  If the DM is a bound state and analogously
$\mu\sim e/2m_\chi$, this gives $\mu \sim 1.1$/TeV, which has the right
order of
magnitude.  This could happen if the hidden sector has a confining gauge
symmetry such as SU(3), analogous to QCD, that becomes strong at the 100 GeV
scale.  However we cannot push the analogy to QCD too far.  Not only do
we need the DM particle to be
absolutely stable, we need it to be the lightest bound state,
since otherwise it would have a large annihilation cross-section into
lighter bound states.  Therefore it cannot be a baryon of
SU$(N_{\rm c}>2)$, since there will always be lighter mesonic states.

There will however be fermionic meson states in theories with both
scalar and spinor matter, say a charged scalar techniquark
$S$ and an oppositely-charged 
anti-techniquark $\psi$ forming a spin-$\frac 12$ bound state
$\eta = S\psi$.
Such models automatically also have composite neutral bosonic mesons
$\tilde\eta_\SSS = S^{*}S$ and $\tilde\eta_\psi =
\bar\psi \psi$.  If the
$\eta$ is lighter than the other mesons and
baryons, it is safe from annihilating into them; and if the $S,\psi$
mass scales are below or comparable to the confinement scale, the
glueballs are also heavier. The details of the strong dynamics may
determine which meson is the lightest, but we argue in the next
section that it can plausibly be the $\eta$.

\begin{figure}[tb]
\centering
\includegraphics[width=0.35\textwidth]{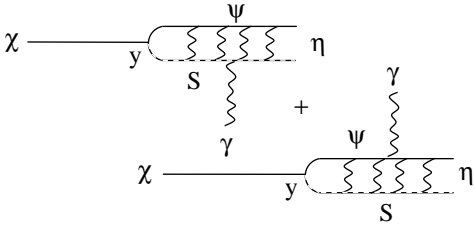}
\caption{Schematic Feynman diagrams for generation of large transition
dipole moment between the elementary state $\chi$ and the composite
one $\eta$.}
\label{figmu}
\end{figure}

A further requirement is to split the Dirac $\eta$ state into Majorana
states of different mass, in order to have a transition dipole moment.
In the theoretically preferred scenario of large
magnetic moment, this splitting has to be sizable, $\gtrsim 10\%$,
to get a sufficient relic density.   A loop effect (such as that which
gives the neutron a mass splitting in $R$-parity violating supersymmetry
\cite{Goity:1994dq,Chang:1996sw}) will be too small.  But there is a simple
way to get a large mass splitting at tree level, if the dark matter is an
admixture of an elementary fermion $\chi$ and the bound state 
$S \psi$, by having a bare
mass term $\frac12 m_{\chi}\bar\chi\chi$ and a Yukawa coupling $y \bar\chi 
\psi S$.  The latter becomes an off-diagonal mass term 
with mass of order  $y\Lambda$ at scales below the confinement scale
$\Lambda$.  Large mass
splittings result as long as $m_{\chi}$ is comparable to $m_\eta$, the
(unmixed) bound state mass.
The generation of the transition moment can be visualized through the
diagrams of fig.\ \ref{figmu}.

Finally, the exotic charged states that appear as bound states (or
within the loop for a perturbative origin of the dipole) must not be stable
since there are very stringent constraints on charged relics.  Thus any model
introducing exotic heavy charged particles to induce the dipole moment
must ensure that
they can decay into charged particles within the standard model, while
respecting the stability of the dark matter. 

\bigskip\noindent
{\bf Explicit model.}  
Based on the above considerations,  we construct the simplest model of
composite dipole dark matter\footnote{For another recent model of
composite DM, without a dipole interaction however, see ref.\
\cite{Alves:2009nf}} that meets all the requirements.   We take the
new confining gauge group to be SU(2)$_g$, and introduce a fundamental
(SU(2)$_g$ doublet) Dirac fermion $\psi$ which
carries electric charge $n{+}\frac 12$ for integer $n$, and a scalar $S$
with the opposite charges. There is an elementary Majorana
fermion $\chi$ that mixes with the neutral $S\psi$ bound state to form
the dark matter.  The particle content and quantum numbers are listed
in table \ref{tab1}.  

\begin{table}[tb]
\begin{tabular}{|c||c||c|c|c||c||c|}
\hline
state & spin & SU(2)$_g$ & U(1)$_y$ & U(1)$_{\rm em}$  & $Z_4$ &
constituents \\
\hline\hline
$\chi$  & $\sfrac12_{\phantom{|}}\!$ & 1 & $\phm$0 & $\phm$0 & $-1$& - \\
$\psi_{a}$ & $\sfrac12$ & $\overline{2}$ & $-(2n{+}1)$ & $-(n+\sfrac12)$ &
$\phm i$ & -\\
$S^a$ & 0 & $2$ & $\phm (2n{+}1)$ & $\phm(n+\sfrac12)_{\phantom{|}}\!$ &  
$\phm i$  & -\\
\hline\hline
$\eta$ & $\sfrac12$ & 1 & $\phm 0$ & $\phm 0$ & $-1$ & $S\psi$ \\
$\tilde\eta_\SSS$ & 0 & 1 & $\phm 0$ & $\phm 0$ & $\phm 1$ & $S^*S$\\
$\tilde\eta_\psi$ & 0 & 1 & $\phm 0$ & $\phm 0$ & $\phm 1$ & $\bar\psi\psi$ \\
\hline
$N^-$ & $\sfrac12$ & 1 & $-(4n{+}2)$ & $-(2n+1)$ & $\phm 1$ & $S^*\psi$\\
$\tilde N^+_\mu$ & 0 & 1 & $\phm (4n{+}2)$ & $\phm (2n+1)$ & $-1$ & $SS$\\
$\tilde N^-_{\psi_{\phantom{|}}}$ & 0 & 1 & $-(4n{+}2)$ & $-(2n+1)$ & $-1$ & $\psi\psi$\\
\hline
\end{tabular}
\caption{Particle content and representations of the confining gauge group,
weak hypercharge, electric charge, discrete 
global DM number symmetry charge, and particle content (in the case of
composite states), for states in the minimal
model of composite dipole dark matter.  Top 3 rows are the elementary
constituents, bottom rows are the mesonic and baryonic bound states.}
\label{tab1}
\end{table}

The relevant mass terms and interactions in the potential at the
scale above $\Lambda$ (the compositeness scale) are 
\bea
V &=&	\sfrac12 m_{\chi}\bar\chi\chi + 
m_\psi \bar\psi\psi + m^2_\SSS |S|^2  +\lambda|S|^4 \\  
	 &+& 
	  \bar\chi\, S^a (y + iy_5\gamma_5)\psi_{a} + 
	y'\epsilon_{ab}S^*_a\bar e_\R \psi_{b}\left(\bar e_\R
e_R^c\over\Lambda'^3\right)^n
 +{\rm h.c.}\nonumber
\label{lag}
\eea
The interactions are invariant under a discrete $Z_4$ symmetry, shown
in table \ref{tab1}, which guarantees the stability of the dark matter.
The last interaction involves electrons for simplicity, but in general
any combination of right-handed charged leptons can appear.

The physical states of the system are the
neutral mesonic bound states $\eta = S^a \psi_a$, 
$\tilde\eta_\SSS = S_a^*S^a$,
$\tilde\eta_\psi = \bar\psi^a\psi_a$, and the charged
baryonic%
\footnote{For SU(2) there is no real distinction between mesons and
  baryons; so our choice to call these baryons is just a convention.}
ones 
$N^- = \epsilon^{ab}S_a^*\psi_b$, 
$\tilde N^+_\mu = \epsilon_{ab}S^a\partial_\mu S^b$,
$\tilde N^-_\psi = \epsilon^{ab}\psi_a\psi_b$.
The state $\eta$ will be the lightest for some part of the parameter
space, since $\tilde\eta_\SSS$ gets a positive mass correction from the
repulsive quartic scalar interaction%
\footnote{There is a potential danger that $\lambda$ runs to negative
  values before the confinement scale, due to a $g^4$ term in its beta
  function $\beta_\lambda$.  This is partly compensated by the $-y^4$
  term in $\beta_\lambda$.  Here we will assume that parameters can be
  found where it does not go negative below the scale $m_S$.}%
, there is a positive spin-spin
interaction energy for $\tilde\eta_\psi$, and we are still free to
choose $m_\chi,m_\SSS$ -- which however cannot be much heavier than the
confinement scale so that $m_\eta < m^{0++}_{\rm glueball}$.
The charged states are heavier due to
positive Coulombic energy shifts.

The bound state $\eta$ mixes with $\chi$ through the first Yukawa 
interaction in (\ref{lag}).  The mass terms relevant for the DM and its
excited states are
\be
	\sfrac12 m_{\chi}\bar\chi\chi + m_y( \bar\chi
	e^{i\theta_y\gamma_5}\eta   +{\rm h.c.})
	+ m_\eta\bar\eta\eta
\ee
where $m_y$ is of order $(y^2+y_5^2)^{1/2}\Lambda$, $\theta_y = \tan^{-1}(y_5/y)$, 
 and $m_\eta\sim\Lambda$ is the mass of the $\eta$
composite fermion.  For simplicity, we impose parity to set
$\theta_y=0$ in the following, though the qualitative features of the
model do not depend upon this choice.  In the basis of Weyl components 
($\eta^c$, $\eta$, $\chi$) 
of a single handedness, this corresponds to a $3\times 3$ Majorana mass
matrix of the form
\be\label{majmass}
	M = \left(\begin{array}{ccc} 0 & m_\eta & m_y\\
			         m_\eta & 0 & m_y\\
				m_y & m_y & m_\chi\end{array}\right)\,.
\ee
As long as $m_y\ne 0$, the mass eigenstates are distinct Majorana
particles, none of which can be paired up into a Dirac particle.  This is
important, since as noted above, a stable Dirac particle with as large a 
direct magnetic moment as $\mu_\eta = 1$/TeV is ruled out by direct detection
constraints.  Notice that in the same basis, the magnetic moment matrix
takes the form
\be
	\mu = \left(\begin{array}{ccc} 0 & \mu_\eta & 0\\
			         -\mu_\eta & 0 & 0\\
				0 & 0 & 0\end{array}\right) \,.
\ee
Thus if $R^\T M R$ diagonalizes the mass matrix, then $R^\T \mu R$ gives
the transition moments in the mass eigenbasis.

In the parity-conserving case, the spectrum of DM states has a simple
analytic form, which we give in appendix \ref{muex}.  The combination
$(\eta-\eta^c)/\sqrt{2}$ does not mix with $\chi$ and retains mass of 
exactly $m_\eta$.  For small $m_y$, the spectrum can be written in the
form
\be \label{masseigen}
	m_i = (m_\eta-\delta m,\ m_\eta,\, m_\chi+\delta m) 
\ee
where $\delta m \cong 2m_y^2/(m_\chi-m_\eta)$.  We argue that
to avoid suppressing $\mu_{12}$ by a small mixing angle, it is
desirable to choose $m_\eta< m_\chi$ so that the lowest two states
are mostly $\eta,\eta_c$.  Then the mixing angle
between the lowest and highest 
states is  $\theta_{13} \cong \sqrt{2} m_y/(m_\chi-m_\eta)$, and the middle
state is purely composite; if $m_\chi<m_\eta$, the masses take the 
inverse hierarchy $m_3<m_2<m_1$.
The transition moments in either case are $\mu_{12}\simeq \cos\theta\mu_\eta
\simeq\mu_\eta$ and $\mu_{23}\simeq \theta\mu_\eta$.  By
analogy to the
nucleons of QCD, we can estimate $\mu_\eta\simeq (n+\sfrac12)e/(2m_\eta/2)\simeq
(2n+1)\times 1.2$/TeV.  Based on the quantitative success of the quark model in
reproducing the baryon magnetic moments \cite{georgi}, this can be
considered as good to $\sim 10\%$, and not just an order of magnitude
estimate.  On the other hand, the process shown in fig.~\ref{fig1}
involves a $\chi_2$ particle which is far off shell, so the hadronic
effective picture is not reliable.  There can be a nontrivial form
factor, as well as other two-photon operators such as
$\bar\chi \gamma^5 \chi F\tilde{F}$.  It is best to consider the
annihilation as one effective operator to two photons, and to adopt the
dipole annihilation calculation, eq.~(\ref{eqB1}),
as a rough estimate for the cross-section which probably receives
${\cal O}(1)$ corrections.  It would be difficult to improve this
estimate analytically, but it is possible that a lattice calculation
could clarify the actual annihilation rate.
We will moreover see in the following section that
it is necessary to take the mass ratio $r=m_{\chi_2}/m_{\chi_1}$ to 
be greater than $2$, hence $f(r)>1.2$ in eq.\ (\ref{edm_range}).
Therefore to achieve a sufficient annihilation rate we may need to
consider $n=1$ (charge-$3/2$ techniquarks), leading to
$\mu_\eta\simeq 3.6/$TeV.

We thus arrive at the favored scenario in which 
$m_\chi > m_\eta > m_y$.  We have computed the ratio of
the transition moment $\mu_{12}$ 
 to its maximum value $\mu_\eta$ as a
function of $\delta m/m_\eta$, which is shown for 
several values of $m_\chi/m_\eta$ in fig.\ \ref{mu12}.  The exact mass 
splittings and mixing angles are given in appendix \ref{muex} in this
parity-symmetric case.  
(We leave the parity-violating case for future work; generally, all three
states will mix, and none of the possible transition moments will vanish.)

\begin{figure}[tb]
\centering
\includegraphics[width=0.4\textwidth]{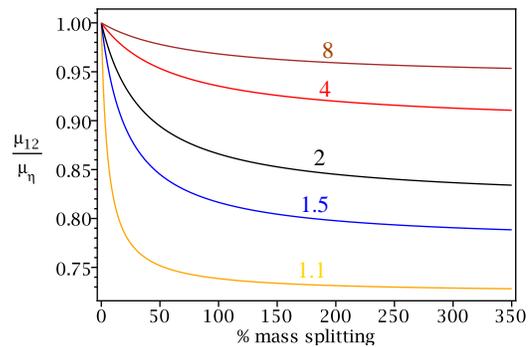}
\caption{Ratio of transition moment $\mu_{12}$ to its maximum value
$\mu_\eta$ as a function of the relative mass splitting between the DM and
its first excited state, for $m_\chi/m_\eta=1.1,\ 1.5,\ 2,\ 4,\ 8$, as labeled.}
\label{mu12}
\end{figure}

The baryonic bound states carry electric charge, and must not be
stable because of very stringent bounds on relic charged particles.  
The $y'$ Yukawa coupling ensures that they can decay into $2n{+}1$
charged leptons and a photon or dark matter.
For the $N^-$ baryon, the $y'$ coupling
in (\ref{lag}) leads directly to a decay
term $\tilde y'(\Lambda/\Lambda'^{3n})\, (\bar e_\R N^-)
(\bar e_\R e_\R^c)^n$, 
Because of their nontrivial $Z_4$ charges, the bosonic
baryons must decay via $\tilde N^+_\mu\to\eta +(2n{+}1)\ell^+$ and
$\tilde N^-_\psi\to \eta +(2n{+}1)\ell^-$ respectively, through the
elementary  processes  $S\to\psi +(2n{+}1)\ell^+$ or
$\psi\to S+ (2n{+}1)\ell^-$.  Clearly, only a
small mass splitting between the baryons and the DM is needed to make
these kinematically possible. We remind the reader that $e_\R$ (hence
$\ell_\pm$) stands for any flavor of right-handed charged leptons.

\begin{figure}[tb]
\centering
\includegraphics[width=0.5\textwidth]{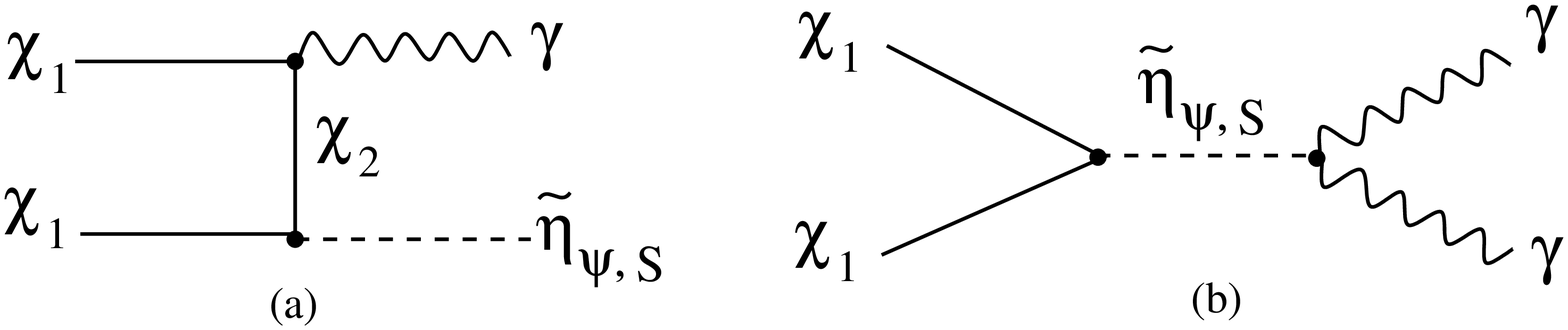}
\caption{New channels for DM annihilation present in the strongly
interacting model, that should not dominate over the magnetic moment
diagram of fig.\ \ref{fig1}(a).}
\label{newrelic}
\end{figure}

\bigskip\noindent
{\bf Challenges to the model.}
Having invented a specific model, we need to check that no important
new annihilation channels were introduced, that were not present when
we assumed the dark matter interacted only through its magnetic moment
coupling.  Due to the strong interactions of the DM constituents,
there are indeed new possible diagrams shown in fig.\ \ref{newrelic}
that can be problematic.  

The first one, fig.\ \ref{newrelic}(a),
illustrates the process $\chi_1\chi_1\to
\tilde\eta_{i}\gamma$  (where $i=\psi,S$) which involves a Yukawa interaction
\be
\tilde y_i\tilde\eta_{i}\bar\chi_1\chi_2
\label{newint}
\ee
 in the effective
theory below the confinement scale.  This vertex only requires the
rearrangment of the constituents into different bound states (with the
annihilation of $SS^*$ or $\psi\bar\psi$), so there is no {\it a
priori} reason for it to be suppressed, and  the couplings $\tilde y_i$
might be relatively large.  Thus this channel will dominate over that
of fig.\ \ref{fig1}(a) if it is kinematically allowed, and one
therefore needs to have $2 m_{\chi_1} < m_{\tilde\eta_i}$.  This can 
be accomplished by judicious choices of the mass matrix elements in
(\ref{majmass}).  For example, with $m_y/\Lambda = 0.55$ and
$m_\chi/\Lambda = 1.5$, we find that
$m_{\chi_3}:m_{\chi_2}:m_{\chi_1} = 2.1:1:0.43$, which is sufficient.
And the transition dipole moments are not much suppressed:
$\mu_{23}/\mu_\eta = 0.6$, $\mu_{12}/\mu_\eta = 0.8$.
The process with off-shell $\tilde\eta_i$ leads to
$\gamma$-ray continuum $\chi_1\chi_1\to 3\gamma$
annihilations, but they are suppressed by a power of $\alpha$ and should
be small.

The second diagram, fig.\ \ref{newrelic}(b), again involves a presumably
large effective Yukawa coupling, and an off-shell $s$-channel
$\tilde\eta_i$ particle.  To appraise it, we need an estimate of
the $\tilde\eta_i$ coupling to two photons. To compute this from first
principles would require knowledge of the nonperturbative matrix
elements, but by comparing to the two-photon decays of the $\eta$ and
$\eta'$ mesons in the real world, we estimate that the interaction takes
the form
\be
	c{q^2 e^2\over\Lambda} \tilde\eta_i F_{\mu\nu}F^{\mu\nu}
\ee
with $c \cong 0.1$, where $q=(2n{+}1)/2$ is the charge of the
constituents in our
model.  The suppression by $c$ is accompanied by an additional
$p$-wave suppression due to the scalar coupling (as opposed to
pseudoscalar) in (\ref{newint}).  The cross section, ignoring
interference with fig.\ \ref{fig1}(a), is
\be
	\sigma = {2 v^2\over \pi} {c^2 \tilde y_i^2 q^4 e^4 m_\chi^4\over
	\Lambda^2 (4m_\chi^2 - m^2_{\tilde\eta_i})^2} \,.
\ee
Therefore this interaction turns out to be subdominant to that of 
fig.\ \ref{fig1}(a) unless it happens to be very close to
resonance.  Note however that the phenomenology of this process is the
same as that of the magnetic moment interaction, in that it also
produces photon pairs which would give a line feature.  Therefore we
might also consider the model where the magnetic moment annihilation
rate is too small, but this process proceeds near threshold and actually
dominates the two-photon production rate.  In this case the model would
be viable with the small ($n=0$) charge assignment for the $S,\psi$
techniquarks.

\bigskip \noindent
{\bf Collider signatures.}  Past studies of the collider
signatures of magnetically interacting dark matter have focused on the
production of the DM and its excited state $q\bar q\to\chi_1\chi_2$,
mediated by $s$-channel $\gamma$ or $Z$ and the transition moment
$\mu_{12}$ of $\chi_1$-$\chi_2$.  The subsequent decay
$\chi_2\to\chi_1\gamma$ through the dipole moment can produce a hard
monophoton  that would pass experimental cuts 
\cite{Chatrchyan:2012tea,atlas}  if the $\chi_1$-$\chi_2$
mass splitting exceeds 125-150 GeV. 
Otherwise the pair-production of DM
can be searched for in a more generic way, through missing energy and
initial state radiation of a hadronic monojet
\cite{Bai:2010hh}-\cite{Fortin:2011hv}.

\begin{figure}[tb]
\centering
\includegraphics[width=0.47\textwidth]{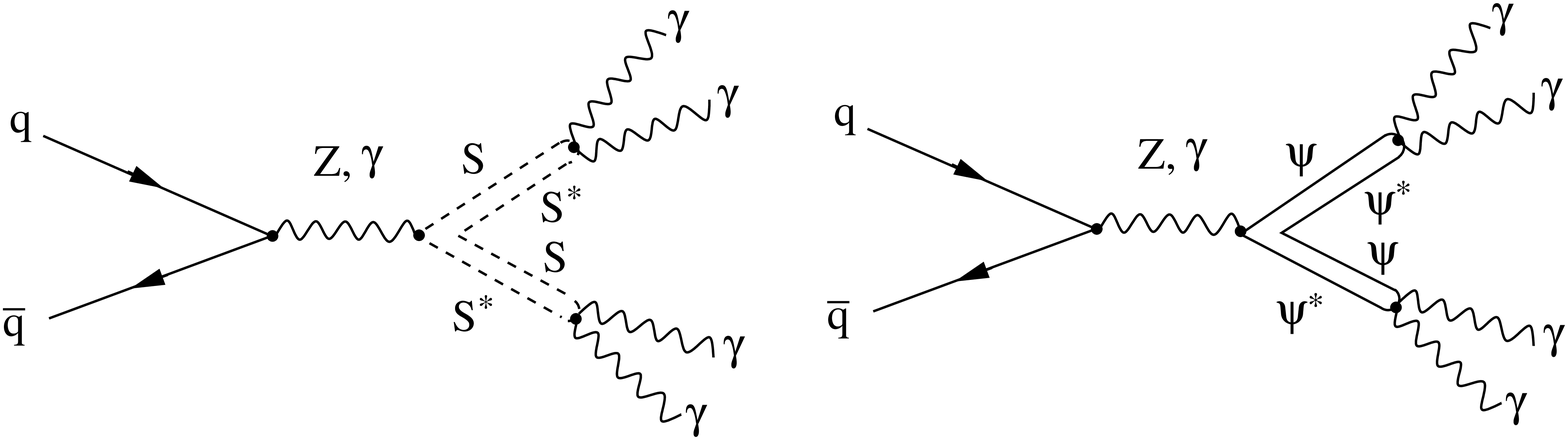}
\caption{Production and decay of $\tilde\eta_\SSS$ and
$\tilde\eta_\psi$ pairs at a hadron
collider.\label{LHC}}
\end{figure}

\begin{figure}[tb]
\centering
\includegraphics[width=0.3\textwidth]{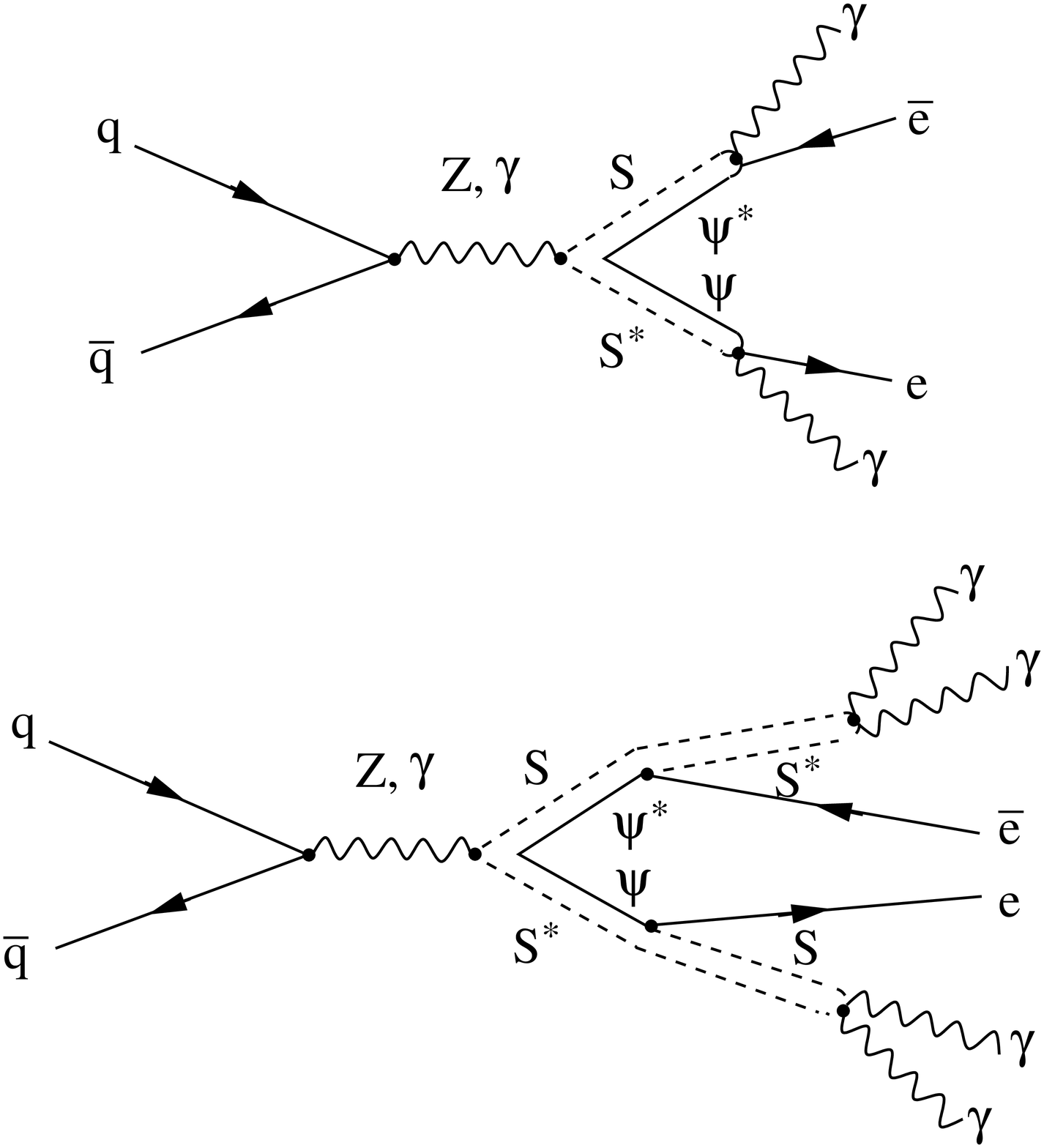}
\caption{Production of $N^\pm$ pairs and two possible decay 
channels (third channel, $e\bar e \, 3\gamma$ not shown).
Here $n=0$ is assumed; for the $n=1$ model, each lepton is replaced by
a triplet of leptons.
\label{LHCb}}
\end{figure}

However the previous considerations assume that the dipole interaction
remains hard at LHC energies.  We have shown that whether one invokes
a loop effect or more plausibly compositeness to generate the large
moment needed to explain the 130 GeV line, it arises from physics
well below the TeV scale, and therefore the magnetic moment effective
description is not valid at high energies.  It will open up to reveal
the constituent particles rather than behaving like a dipole moment.
Equation (\ref{sig_est}) below estimates the cross section for dark 
matter production (though it is likely further suppressed due to 
cancellation between the oppositely charged DM consituents and slightly
due to the DM mixing angles).

In the model we have proposed, the primary process will be to produce
$\psi\bar\psi$ and $SS^*$ pairs via virtual photon and $Z$.  These
will then hadronize into the mesons and baryons listed in table
\ref{tab1}.  The final states will include
the $\chi_1$ and $\chi_2$ (mostly $\eta$) dark matter particles,
producing monophotons and monojets.
But other composite states will also be produced, and they can
have very different signatures because of their visible decays.  For
example, the bosonic bound states $\tilde\eta_\SSS$ and
$\tilde\eta_\psi$ can decay into two photons by direct annihilation of
their constituents, as shown in fig.\ \ref{LHC}.  Thus a very clean
signal of two diphoton pairs,
each of which has the same invariant mass $\gtrsim 260$ GeV, is
predicted.  The production of these particles is suppressed near
threshold because of the overall charge neutrality of the mesons,
leading to {\it e.g.}, destructive interference in the photon+$Z$
coupling to $\psi$ and to $\bar\psi$.  But at center-of-mass energies
well above the particle masses, the
vector boson resolves the constituents and couples fully to both of
them.  The spin and color summed/averaged partonic cross section for
$\psi\bar\psi$ production from up quarks, at leading order in $1/s$, is
given by%
\footnote{$4/9$ and $1/36$ are the squared hypercharges of the right and
  left components of the up quark.}
\be
	\sigma \cong {4\pi\alpha^2 Q_\psi^2 \over 9s \cos^4\theta_\W}
 \left( \frac{4}{9}+\frac{1}{36} \right) \quad 	\sim \quad 6{\rm\ fb}
\label{sig_est}
\ee	
where for the numerical estimate we take $Q_\psi=1/2$ and $\sqrt s=1$ TeV.
Of course a serious prediction requires weighting by parton 
distribution functions, which will reduce (\ref{sig_est}) by around an
order of magnitude because of
the need to get the antiquark out of the sea, as well as a better
treatment of the hadronization process.
But this estimate already shows that the production is at a potentially
interesting level for discovery at the LHC; indeed, since the four
photon signal is so striking and involves particles of such high energy,
the existing LHC data may already have sufficient sensitivity to exclude
or discover this model.

Even more interesting perhaps is the pair production of charged
technibaryons, especially the $N^-$, which can decay via $N^-\to
(2n{+}1)e+\gamma$, or $N^-\to (2n{+}1)e+\tilde\eta_{\SSS,\psi}$
if the latter is
kinematically allowed.  Unlike the neutral pairs, here the production
is not suppressed at threshold because  both constituents have charges
of the same sign.   If both channels are open, the possible final
states are $(2n{+}1)e+(2n{+}1)\bar e$ plus two, three or four photons,
due to the subsequent decays of the $\tilde\eta_{\SSS,\psi}$ mesons. This is
illustrated in fig.\ \ref{LHCb}.  However it is quite possible that
the $N^-\to (2n{+}1)e+\tilde\eta_{\SSS,\psi}$ decays are kinematically
blocked, in which case only the
$(2n{+}1)e\,(2n{+}1)\bar e\,\gamma\gamma$ channel is
present.  Moreover, there is no reason to suppose that the $y'$
coupling in (\ref{lag}) is only to electrons, and it may be natural to
assume that the dominant coupling is to $\tau$ leptons, which would be
more difficult to identify in the final state.

Finally, there will be pair production of the charged bosonic
technibaryons $\tilde N^\pm_\mu$ and $\tilde N^\pm_\psi$, which decay
into leptons and dark matter.  This however has a large standard model
background (if $n=0$) since such events could be mistaken for $W$ boson
decays.

\bigskip {\bf Conclusions.}  
If the 130 GeV gamma ray line and the tentative accompanying line at
114 GeV  are indeed due to dark matter annihilation into
$\gamma\gamma$ and $\gamma Z$,  it is a challenge to explain its
relatively strong annihilation into two monoenergetic photons, while
keeping its annihilation into other particles that produce a continuum
of secondary photons within bounds.   We observed that a DM species
comprising just 10-15\% of the total DM mass  density, possessing a
transition magnetic moment of $\sim 2$/TeV and a mass splitting $\gtrsim
10$ GeV can satisfy these requirements, with its relic density
determined by the annihilations into photons and $Z$ bosons.  (Although
an electric dipole moment of the same strength and smaller mass
splitting could do the same, and be relevant for direct detection, we
find it theoretically difficult to generate such a large EDM without
an accompanying MDM of the same size.)   This scenario leaves open the
possibility of direct detection of some other, dominant DM species.

We further observed that it is difficult to explain a dipole moment of
the needed magnitude through a loop effect (see beginning
of ``Model requirements'' section), which motivated us to study a
composite DM particle.  We presented a simple model
that accomplishes this with an electrically charged ``quark'' and  
scalar ``quark'' of a new SU(2) gauge interaction that confines at the
100 GeV scale, and whose neutral mesonic bound state mixes with an
elementary fermion to make the dark matter and its excited state. 
Remarkably, the magnetic moment is determined by the charge and mass
of the constituents, leaving relatively little room for adjustment
given that the DM has mass 130 GeV, yet coming out to approximately
the desired value.  Unfortunately, complicated confining dynamics
occur at the scale of the annihilation momentum transfer, so the
cross-section can have an ${\cal O}(1)$ difference from the pure
dipole calculation.

Moreover the model makes an unambigous prediction for hadron collider
production  of pairs of ``unflavored'' scalar-scalar or
fermion-fermion bound states of mass $\gtrsim 200$ GeV, whose only
decay channel is into two photons, and whose production cross section
is estimated to be within reach of the LHC.  This double diphoton 
pair signal comes in addition to the more discussed signature of DM
pair production with missing energy and monophotons or monojets.  And due to the pair
production of charged bound states, further exotic final states can
appear where two photons and one lepton (or three leptons in a
related model) have the invariant mass of the charged parent.  Although there
is some freedom in varying the details that accommodate the decays of the
charged bound states (necessary in order to avoid highly constrained
charged relics), the two-photon decay of the neutral bound states
is a robust prediction in any model that uses compositeness
to generate large magnetic moments.  

The dark matter interpretation of the 130 GeV line is expected to be
tested definitively by upcoming gamma ray experiments
\cite{Bergstrom:2012vd}, in particular by HESS-II which begins very
soon.  It is exciting to consider that  complementary indirect
evidence could come from the LHC on a similar time scale, if a new
peak at mass $\gtrsim 260$ GeV (according to the discussion below
eq.\ (\ref{newint})), coming from pairs of 
diphoton events, should be discovered.

\bigskip
 {\bf Acknowlegments.}  We thank 
M.\ Cirelli, M.\ L\"uscher, V.\ Sanz, G.\ Servant, J.\ Shigemitsu, R.\ Teuscher, M.\ Trott,
and C.\ Weniger for helpful discussions or communication.
JC thanks the CERN theory division for its kind hospitality during the
commencement of this work.
This work was supported in part by the Natural Sciences and Engineering
Research Council of Canada (NSERC).

\appendix

\section{$\chi_1\chi_2\to f\bar f, WW, hZ$ coannihilation}
\label{coann}
The annihilation cross section for $\chi_1\chi_2\to f\bar f$ for
a standard model fermion pair at low velocity, via a magnetic moment
interaction, can be expressed as
\be
	\sigma_{\bar f f} v = \alpha\mu^2 \sum_i N_i
\ee
where $N_i=1$ for a hypothetical particle with unit electric
charge and no coupling to the $Z$.   The sum is over all kinematically
accessible final states.   For standard model fermions,
and similarly parametrizing the contributions from  
$W^+W^-$ and $Zh$ final states, $N_{\rm eff}$ is
\be
	N_i = \left\{ \begin{array}{ll} q_i^2(1-2v_it_\W\xi
	+(v_i^2+a_i^2)t_\W^2\xi^2)), & f \bar f\\
	{1\over 16}\left(m_Z\over m_W\right)^4
	\xi^2\,\psi_{\W\W},& WW\\
	{1\over 16\, c_\W^4}\,\xi^2\,\psi_{hZ},& hZ
\end{array}\right.
\ee
The $\bar f f$ result is as given in ref.\ \cite{Weiner:2012cb},
where $q_i$ is the electric charge, $v_i,a_i$ are the vector and
axial-vector couplings respectively to the $Z$, in units of
$q_i e$, and $\xi = (1-m^2_Z/4(m^2_\chi))^{-1}$.  For
$WW$ we define $\psi_{\W\W} = \left(1 + 4\epsilon_\W - {17\over 4}
\epsilon_\W^2 -	\frac34\epsilon_\W^3\right)(1-\epsilon_\W)^{1/2}$
with $\epsilon_\W = m_W^2/m_\chi^2$.
We find that
the contribution from neutrinos is $\sum_{i} N_i =
3\xi^2/(8c_\W^4)$, and for all charged leptons or quarks
 $v_i=t_\W((4s_\W^2|q_i|)^{-1}-1)$, $a_i = t_\W/(4s_\W^2|q_i|)$,
giving $\sum_i N_i = 10.4$ from fermionic final states, assuming 
$m_\chi = 130$ GeV and $s_\W^2 = 0.23$.  Because of a strong
cancellation between the virtual $\gamma$ and $Z$ contributions to 
the $WW$ channel, it contributes only $0.20$ to $\sum_i N_i$.
For $hZ$, we define $\psi_{hZ} = \left(1-\frac12(\epsilon_h - 5\epsilon_\Z)
+\frac{1}{16}(\epsilon_h-\epsilon_\Z)^2\right)\times$ $\left(1-\frac12(\epsilon_h+\epsilon_\Z)
+ \frac{1}{16}(\epsilon_h-\epsilon_\Z)^2\right)^{1/2}$ with
$\epsilon_h = m_h^2/m_\chi^2$, contributing $0.22$ to $\sum_i N_i$.
Hence the total is $\sum_i N_i = 10.8$.

For the $f\bar f$ channels, the result 
from an electric dipole moment interaction is the same, except for
 an additional velocity-squared suppression factor 
of $v^2/3$.

\section{$\chi_1\chi_1\to\gamma\gamma$, $\gamma Z$, $ZZ$ annihilation}
\label{ann}
Defining $r=m_{\chi_2}/m_{\chi_1}$ and $\epsilon_\Z = \frac14(m_Z/m_{\chi_1})^2$
(superseding the definition of $\epsilon_\Z$ in appendix \ref{coann}),
the cross sections (times relative velocity) for $\chi_1\chi_1$ to annihilate into $\gamma\gamma$, $\gamma Z$
and $ZZ$ are given by
\bea
	\langle\sigma v\rangle = {\mu^4 m^2_{\chi_1}\over 4\pi}\left\{
	\begin{array}{ll} {4r^2\over (1+r^2)^2},& \gamma\gamma\\
	2 t_\W^2 {(1-\epsilon_\Z)^3(r+\epsilon_\Z)^2\over
	\left(\sfrac12(1+r^2)-\epsilon_\Z\right)^2},& \gamma Z\\
	t_\W^4{(1-4\epsilon_\Z)^{3/2} (r+2\epsilon_\Z)^2\over
	\left((\sfrac12(1+r^2)-2\epsilon_\Z\right)^2},& Z Z
	\end{array}\right.
\label{eqB1}
\eea
The Dirac case corresponds to $r=1$.
\bigskip 

\section{Explicit $3\times 3$ DM mixing.}
\label{muex}

The parity symmetric $3\times 3$ Majorana mass matrix $M$ given in 
(\ref{majmass}) can be diagonalized exactly, by converting it to a block
diagonal  $1\times 1 \oplus 2\times 2$ matrix. 
The exact mass eigenvalues are $-m_\eta$, which corresponds to an
eigenvector entirely in the subspace of composite particles, and
\be 
\frac{1}{2}\left[m_\chi+m_\eta \pm \sqrt{(m_\chi-m_\eta)^2+8 m_y^2}\,\right]
\ee
with eigenvectors mixing $\chi$ and the composite states.  The physical
masses are then as in equation (\ref{masseigen}) with
\be 
\delta m\equiv \frac{1}{2}(m_\eta-m_\chi)\left[1 - 
\sqrt{1+8 m_y^2/(m_\chi-m_\eta)^2}\,\right]\ .
\ee
Notice that $\delta m > 0$ when $m_\chi> m_\eta$, leading to level
repulsion.

We can now write
the three eigenvectors as $\chi_1=\cos\theta\, \eta_2+\sin\theta\,\chi$,
$\chi_2=\eta_1$, and $\chi_3=\cos\theta\,\chi-\sin\theta\,\eta_2$ 
in order of increasing
mass and promote them to 4-component Majorana spinors.  Then the 
transition dipole moment between the two composite states becomes
\be {\mu_\eta} \bar\eta_1\sigma_{\mu\nu}\eta_2F^{\mu\nu}
= {\mu_\eta} \bar\chi_2\sigma_{\mu\nu}(\cos\theta\,\chi_1-\sin\theta\,
\chi_3)F^{\mu\nu}\ .\ee
The transition moment between the two lightest states is therefore
suppressed by the cosine of the mixing angle $\theta$, and there is no
transition moment between the lightest and heaviest states.

Finally, explicitly finding the eigenvectors yields a mixing angle
\be \cos\theta = 2m_y/\sqrt{2\delta m^2 +4m_y^2}\ .\ee
In terms of $m_\chi,m_\eta,\delta m$, this becomes
\be \cos\theta =\sqrt{\frac{\delta m+m_\chi-m_\eta}{2\delta m+m_\chi-m_\eta}}
\ .\ee
A straightforward further algebraic rearrangement is necessary to 
write this in terms of $m_\chi/m_\eta$ and the relative 
mass splitting $\delta m/(m_\eta-\delta m)$ as in figure \ref{mu12}.

\bigskip	 


\bibliographystyle{apsrev}

\end{document}